\documentclass[journal]{IEEEtran}

\usepackage{amsmath}
\usepackage{epsfig}
\usepackage{epstopdf}
\usepackage{subfig}
\usepackage{graphicx}
\usepackage{float}
\usepackage{algorithm,algorithmic}
\usepackage{url}
\usepackage{color}
\usepackage{setspace}
\def\tcr{\textcolor{black}}

\ifCLASSINFOpdf
\else
\fi

\begin{document}
	\title{Prior Information Guided Regularized Deep Learning for Cell Nucleus Detection}
	
	\author{Mohammad~Tofighi,~\IEEEmembership{Student Member,~IEEE,}
		Tiantong~Guo,~\IEEEmembership{Student Member,~IEEE,}
		Jairam~K.P.~Vanamala,
		and~Vishal~Monga,~\IEEEmembership{Senior Member,~IEEE}% <-this % stops a space
		\thanks{M. Tofighi, T. Guo, and V. Monga are with the Department
			of Electrical Engineering, Pennsylvania State University, University Park, PA. e-mail: tofighi@psu.edu, tiantong@psu.edu, juv4@psu.edu, vmonga@engr.psu.edu.}% <-this % stops a space
		\thanks{J.K.P. Vanamala is with Center for Molecular Immunology and Infectious Disease, Pennsylvania State University, University Park, PA.}
		\thanks{Research has been supported by an NSF CAREER award (to V. Monga).}}

	\markboth{IEEE Transactions on Medical Imaging, Accepted for Publication, January 2019}%
	{Shell \MakeLowercase{\textit{et al.}}: Bare Demo of IEEEtran.cls for IEEE Journals}

	\maketitle
	\begin{abstract}
		Cell nuclei detection is a challenging research topic because of limitations in cellular image quality and diversity of nuclear morphology, i.e. varying nuclei shapes, sizes, and overlaps between multiple cell nuclei. This has been a topic of enduring interest with promising recent success shown by deep learning methods. These methods train Convolutional Neural Networks (CNNs) with a training set of input images and known, labeled nuclei locations. Many such methods are supplemented by spatial or morphological processing. Using a set of canonical cell nuclei shapes, prepared with the help of a domain expert, we develop a new approach that we call Shape Priors with Convolutional Neural Networks (SP-CNN). We further extend the network to introduce a shape prior (SP) layer and then allowing it to become trainable (i.e. optimizable). We call this network tunable SP-CNN (TSP-CNN). In summary, we present new network structures that can incorporate `expected behavior' of nucleus shapes via two components: {\em learnable} layers that perform the nucleus detection and a {\em fixed} processing part that guides the learning with prior information. Analytically, we formulate two new regularization terms that are targeted at: 1) learning the shapes, 2) \tcr{reducing} false positives while simultaneously encouraging detection inside the cell nucleus boundary. Experimental results on two challenging datasets reveal that the proposed SP-CNN and TSP-CNN can outperform  state-of-the-art alternatives.
	\end{abstract}
	
	
	\IEEEpeerreviewmaketitle

	\section{Introduction}
	\label{sec:intro}
	
	%Detection of cell nuclei in microscopic images is to approximately locate the centroid of the cell nuclei.
	Microscopic images often exhibit high degree of cellular heterogeneity \cite{al2003prospective,prince2007identification}.  Cell nucleus detection methods locate the cell and annotate the center of the cell nuclei. Visual-based techniques in many medical imaging applications, e.g. manual detection and counting, are extremely time consuming and do not scale well as the task is to be performed for a large number of images \cite{costes2004automatic,schindelin2012fiji}. Therefore, automatic analysis of cellular imagery to determine nuclei locations is a centrally important problem in diagnosis of several medical conditions including tumor and cancer detection \cite{SC_CNN,mitchell2008circulating}. 	
	
	\noindent \textbf{Related work:}
	Some of the earliest attempts at cell nuclei detection involved tailored feature extraction and morphological processing \cite{LIPSyM,veta2014breast,al2010improved,ali2012integrated}. One key limitation of these approaches is that the best features for nuclei detection are rarely readily apparent. Further, often the designed techniques are too specific to the choice of dataset and not versatile. Because of their ability to perform feature discovery and generalizable inference, deep learning methods have recently become popular for this problem  \cite{shen2017deep,litjens2017survey,xie2018efficient,ram2018joint}. For instance, Cruz-Roa \textit{et al.} \cite{Cruz-Roa2013} showed that a deep learning architecture for nuclei detection outperforms methods based on different image representation strategies e.g. bag of features, canonical and wavelet transforms. Xie \textit{et al.} \cite{Xie2015}, proposed a structural regression model for CNN, where a cell nuclei center is detected if it has the maximum value in the proximity map. In \cite{Xu2016}, Xu \textit{et al.} proposed a cell detection method based on a stacked sparse autoencoder, where it learns high level features of cell centroids and then a softmax classifier is used to separate the nuclear and non-nuclear image patches. Sirinukunwattana \textit{et al.} proposed SC-CNN \cite{SC_CNN} which uses a regression approach to find the likelihood of a pixel being the center of a nucleus. In SC-CNN, the probability values are topologically constrained in a way that in vicinity of nuclei center the probability is higher. Another recent approach \cite{Xing2016} uses a combination of well-known traditional CNNs for cell nuclei segmentation and dictionary learning techniques for refining segmentations results. Combining CNNs with sparse coding, \cite{CSP_CNN} uses a sparse convolutional autoencoder (CSP-CNN) for simultaneous nucleus detection and feature extraction. In CSP-CNN, the detection is based on a fairly deep network ($>$ 15 layers) comprising of multiple CNN branches that perform detection, segmentation and  image reconstruction.
	
	We develop a  novel prior guided learning approach by exploiting a prior understanding of the shape of the nuclei, which can be obtained in consultation with a domain expert. The value of shape based information has been recognized for medical image segmentation \cite{song2017accurate,pan2017accurate}. Leventon \textit{et al.} \cite{Leventon_SP_2000} proposed a method for medical image segmentation by incorporating shape information into the geodesic active contour method. Ali \textit{et al.} \cite{Ali_SP_2012} incorporated prior shape information into boundary and region based active contours for accurate segmentation of cells. In \cite{oktay2018anatomically}, Oktay \textit{et al.} propose to incorporate shape prior information into enhancement and segmentation of MR cardiac images. They propose a fidelity term between the output segmentation regions and the ground-truth region shapes. While the approach in \cite{oktay2018anatomically} also exploits anatomical shape information, technically the proposed SP-CNN is fundamentally different. First, the image modality in our work is cellular as opposed to organ (MR) imagery in \cite{oktay2018anatomically}, which leads to vastly different quantitative formulations. Second, our proposed regularizer is not a fidelity term but guides cell nuclei detection in regions where the probability of presence of a cell nuclei is higher according to the shape prior knowledge. Finally, architecturally the proposed SP/TSP-CNN and the neural network in \cite{oktay2018anatomically} are entirely different. 

	Prior information with deep networks has emerged as a promising direction in imaging inverse problems such as super-resolution \cite{mousavi2018deep, cherukuri2018deep}. These methods use output image priors to enhance the super-resolution task. Similarly in related work, a Deep Image Prior \cite{ulyanov2017deep} network has been specifically designed for tasks such as image in-painting, de-noising, and super-resolution. Shape based priors in a deep learning based nuclei detection framework however remain elusive and form the focus of our work.
	
	\noindent \textbf{Contributions:}
	While existing deep learning approaches for cell nuclei detection are promising, our goal is to fundamentally alter the learning of the network by enriching it with domain knowledge provided by a medical expert. Specifically, our key contributions\footnote{A preliminary 4 page version of this work was published at IEEE ICIP 2018 in Athens, Greece  held in October 2018 \cite{SPCNN}.} are as follows:
	
	\begin{itemize}
		\item \textbf{Shape Prior with a Convolutional Neural Network (SP-CNN):} We propose a novel network structure that can incorporate `expected behavior' of nuclei shapes via two components: {\em learnable} layers that perform the nucleus detection and a {\em fixed} processing part that guides the learning with prior information. Three sources contribute to generating the prior: network output, raw edge map from the input image, and a set of predefined shapes prepared by the medical expert. These shapes are binary images representing the boundary of cell nuclei. The {\em fixed} priors guide the {\em learnable} layers to perform detection consistent with a nucleus boundary.
		
		\item \textbf{Tunable Shape Priors with a Convolutional Neural Network (TSP-CNN):} As a key extension, we develop a new approach where the set of shapes is represented as a convolutional layer, which is no-longer fixed but learned. That is, using the expert provided shapes as a starting point, we both refine and remove redundancy from the shape set (using shape similarity measures) to arrive at a new learned shape set that is not only more economical but also enhances detection accuracy.
		
		\item \textbf{Novel Regularization Terms:} Analytically, we formulate two new regularization terms: a shape prior term and a shape learning term. The shape  prior term is used in both SP-CNN and TSP-CNN and incorporates the effect of prior information by using a set of shapes to guide the learning of the network towards greater accuracy. The shape learning term is only used in TSP-CNN and achieves the refinement/learning of shapes in a constrained manner, i.e.\ by emphasizing similarity to a reference shape set. We carefully design these terms so they are differentiable w.r.t. the output and hence the network parameters, enabling tractable learning through standard back-propagation schemes.
		
		\item \textbf{A New Dataset, Broad Experimental Validation and Reproducibility:} Experimental validation of SP-CNN and TSP-CNN is carried out on two diverse cancer tissue datasets to show its broad applicability. We also provide a new open source expert annotated dataset of microscopic colon tissue images for cell nuclei detection. We call it the PSU Dataset and it is prepared by the help of experts in Center for Molecular Immunology and Infectious Disease, Penn State University \cite{webpage}. The second dataset is courtesy of Sirinukunwattana \textit{et al.} at Department of Computer Science, University of Warwick (we call it UW Dataset) which includes colorectal adenocarcinoma images \cite{SC_CNN}. Extensive experimental results in the form of benchmark measures such as F1 scores and precision-recall curves show that SP-CNN and TSP-CNN outperform many recent methods \cite{SC_CNN,Xie2015,LIPSyM,Xu2016,CRImage,CSP_CNN}, particularly in very challenging scenarios. We also make our code and the PSU dataset freely available at our project webpage \cite{webpage}.
		
	\end{itemize}
	
	The rest of this paper is organized as follows. The proposed SP-CNN structure that incorporates shape priors into cell nuclei detection and formulation of the new regularized cost function are detailed in Section \ref{sec:SPCNN}. Section \ref{sec:TSPCNN} presents the extension of SP-CNN to tunable shape priors. Detailed experimental comparisons against competing state-of-the-art methods are provided in Section \ref{sec:experiments}. Section \ref{sec:conclusion} summarizes the findings and concludes the paper.
	
	\begin{figure}[t!]
		\begin{center}
			\includegraphics[width=85mm, trim={0 0 0 0},clip]{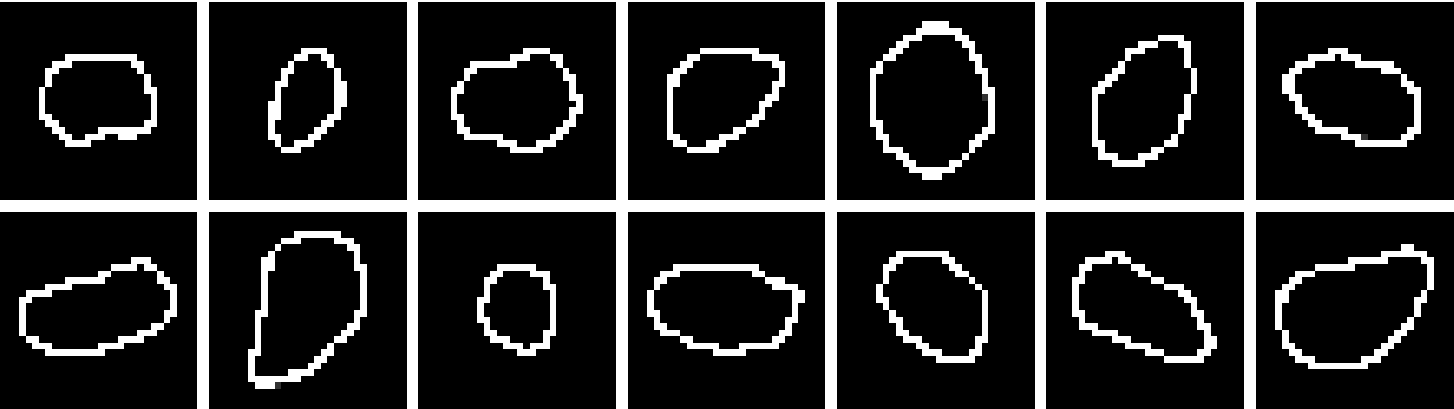}
			\caption{Samples of handcrafted cell nuclei shapes from colon tissues in UW Dataset. For dataset details see Sec. \ref{sec:experiments}.}
			\label{fig:shapes}
		\end{center}
	\end{figure}
	
	\section{Shape Priors with Convolutional Neural Networks (SP-CNN)}
	\label{sec:SPCNN}
	
	\subsection{CNN for Nucleus Detection}
	\label{sec:cnd}
	SP-CNN detects the cell nucleus using a regression CNN. In regression networks, the goal is to obtain a function that interprets the relationship between the input image $\mathbf{x}$ and ground-truth labeled image $\mathbf{y}$. The network is modeled by parameters set $\mathbf{\Theta} = \{\mathbf{W}_l, \mathbf{b}_l\}_{l=1}^{D}$, where $\mathbf{W}_l$ and $\mathbf{b}_l$ denote respectively the weights and bias of $l$-th layer of total $D$ layers collectively. The CNN is learned by solving the well-known optimization problem \cite{SC_CNN,RegFuncRef1}:
	\begin{equation}
	\mathbf{\Theta} = \arg\min\limits_{\mathbf{\Theta}}\|f(\mathbf{x}; \mathbf{\Theta}) - \mathbf{y}\|_2^2
	\label{eq:CNN_func}
	\end{equation}
	where $f(\mathbf{x;\Theta})$ represents the non-linear mapping of the CNN that generates the detection maps $\mathbf{\hat{y}}$. We work with soft labels $\mathbf{y}$ and $\mathbf{\hat{y}}$, i.e.\ $\mathbf{y}, \mathbf{\hat{y}}$ can take values in the range $[0,1]$. In practice, $\mathbf{y}$ is obtained by processing the binary image (0 or 1 at each pixel) of ground-truth nuclei locations as in \cite{SC_CNN}; also see Section \ref{sec:experiments}. Each of the $D$ CNN layers comprises of a convolutional layer followed by an activation function, which is a Rectified Linear Unit (ReLU) \cite{relu} (except the last layer).
	
	\begin{figure*}[ht!]
		\centering
		\includegraphics[width=0.95\linewidth]{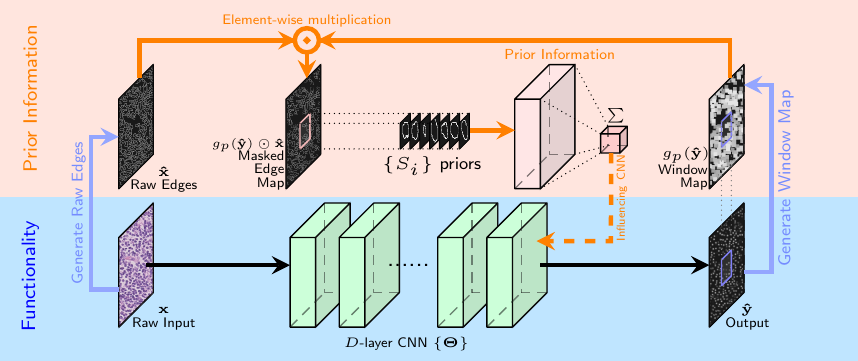}
		\caption{SP-CNN illustration. There are two parts of the SP-CNN: Functionality part (blue) and Prior Information part (orange). The Functionality part consists of one $D$-layer CNN that takes input raw image $\mathbf{x}$ and generates the detected labels for cell nuclei $\mathbf{\hat{y}}$. The prior information part computes the prior cost term as in Eq. (\ref{eq:sp}) and feeds the information into the (learning of the) CNN through back-propagation to guide it towards enhanced nuclei detection. Note the prior information part generates a regularization term used in training the $D$ learnable layers of the CNN and {\em only} the functionality part of the network is applied to a test image. Hence, while the black, orange, and blue arrows are used during training, in testing stage only the black arrows are used.\vspace{-7pt}}
		\label{fig:diagram}
	\end{figure*}
	
	\subsection{Deep Networks With Shape Priors}
	As discussed in Section \ref{sec:intro}, we now incorporate the prior information about the shape of the cell nuclei into the training of the CNN. Ideally, the labels produced by the CNN should lie inside of the nuclei boundaries. We set up a regularization term to explicitly encourage the learned network to achieve detection inside the nucleus boundary while simultaneously \tcr{reducing} false positives. The regularizer is based on a set of shape priors developed with the help of domain expert and given by:
	$\mathbf{S} = \{\mathcal{S}_i | i = 1, 2, \dots, N\}$.
	
\begin{figure}
		\centering
		\includegraphics[width=0.5\textwidth]{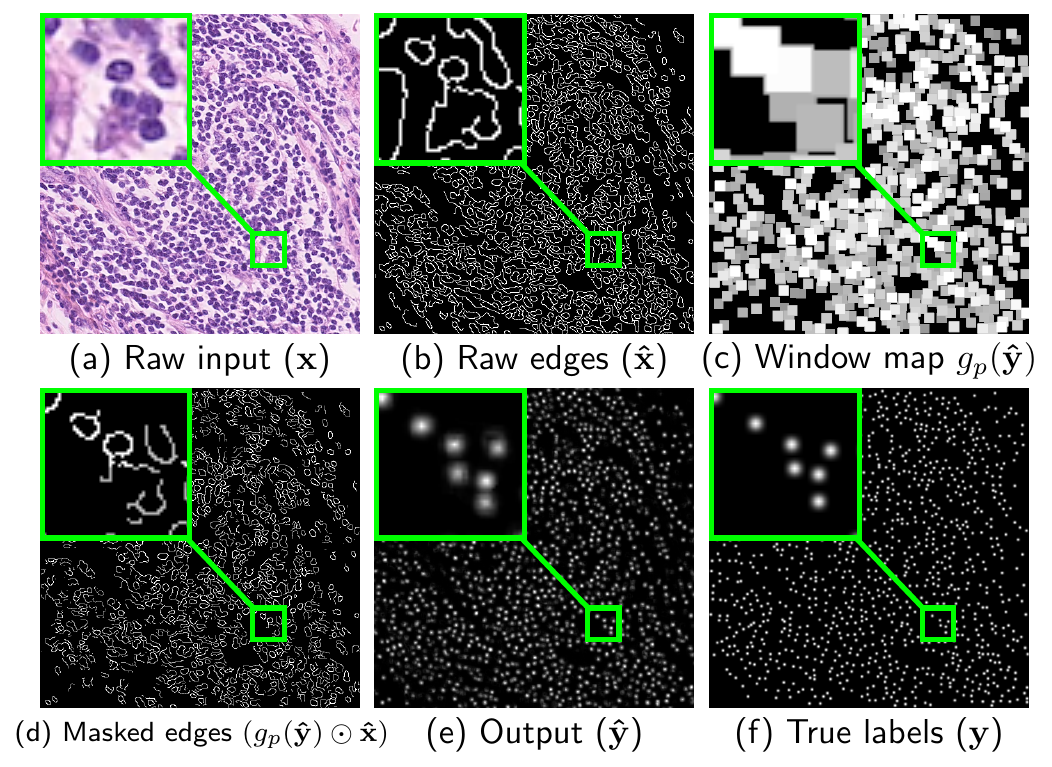}
		\caption{Images in each step of SP-CNN.\vspace{-7pt}}	\label{fig:sample_patches}
	\end{figure}

	For each dataset, multiple training images are analyzed by a medical expert to hand label the nuclei boundaries. A set of $N$ representative shapes is then hand crafted by the medical expert to form the set $\mathbf S$. Some examples of the nucleus shape priors are shown in Fig. \ref{fig:shapes}. These are corresponding to colon tissue images -- detailed explanation is provided in Section \ref{sec:experiments}.  To construct a meaningful regularization term emphasizing shape priors, we need the nucleus boundary information of the input raw image $\mathbf x$. We employ the widely used Canny edge detection\footnote{Note we select Canny edge detection filter because of its simplicity and efficiency. Although its performance satisfies our intentions, other edge detection methods can also be used.} filter \cite{Canny} to generate the raw edge image $\mathbf{\hat{x}}$ with edges labeled as $1$ and background as $0$, as shown in Fig. \ref{fig:sample_patches}-(b). Note that the raw edge image $\mathbf{\hat{x}}$ is only used during the training process.
	We now define the regularization term that captures shape priors:
	\begin{equation}
	\mathcal{L}^{\text{SP}} = -\lambda \sum\limits_{i=1}^{N} \|(g_{p}(\mathbf{\hat{y}})\odot \mathbf{\hat{x}}) \ast \mathcal{S}_i \|_2^2\label{eq:sp}
	\end{equation}
	where the term $g_{p}(\cdot)$ denotes the max pooling operation on $\mathbf{\hat{y}}$ with window size $p$, $\odot$ represents element-wise multiplication, and $\ast$ is the 2D convolution operation. Based on Eq. (\ref{eq:sp}), the computation of the shape priors cost term consists of three steps as shown in Fig. \ref{fig:diagram}'s prior information (orange) part: 
	\begin{enumerate}
		\item The CNN output $\mathbf{\hat{y}}$ is first thresholded by $T_p = 0.2$ to eliminate the background noise (Fig. \ref{fig:sample_patches}-(e)) and then max pooled by $g_p(\cdot)$ with stride of 1 and the `SAME' padding scheme\footnote{Because there is no change in the image grid, this is similar to the box-filtering operation in image processing.}. This results in a window map $g_p(\mathbf{\hat{y}})$ that has $p\times p$ window centered at each location within the soft detected region (Fig. \ref{fig:sample_patches}-(c)). 
		%which has a constant valued $p\times p$ window at each detected location. 
		As expected, a window with higher numerical value will result if the detected label values (in $\mathbf{\hat{y}}$) are correspondingly higher (closer to 1).
		\item The window map $g_p(\mathbf{\hat{y}})$ is then multiplied with the raw edge image $\mathbf{\hat{x}}$ element-wise. This step serves to mask out the edges from $\mathbf{\hat{x}}$ that surround the detected location in $\mathbf{\hat{y}}$, as shown in Fig. \ref{fig:sample_patches}-(d). 
		\item The masked edge image $(g_{p}(\mathbf{\hat{y}})\odot \mathbf{\hat{x}})$ is convolved with the shape priors in set $\mathbf{S}$ to generate a measurement of how well does the detection fit inside the nucleus shape. If $\mathbf{\hat{y}}$ has more labels predicted inside the nucleus boundary, Eq. (\ref{eq:sp}) will achieve a smaller value (more negative).
	\end{enumerate}

	Note that the effect of the shape prior is captured by a negative regularization term since the goal is to maximize (and not minimize) correlation with `expected shapes'. Overall, the cost function of the SP-CNN is given by:
	\begin{align}
	\mathbf{\Theta} &= \arg\min\limits_{\mathbf{\Theta}}\mathcal{L}^{\text{Loss}} + \mathcal{L}^{\text{SP}} \label{eq:costFunc}\\
	&= \arg\min\limits_{\mathbf{\Theta}}\|f(\mathbf{x}; \mathbf{\Theta}) - \mathbf{y}\|_2^2 - \lambda\sum\limits_{i=1}^{N}  \|(g_{p}(\mathbf{\hat{y}})\odot \mathbf{\hat{x}}) \ast \mathcal{S}_i \|_2^2\nonumber{}
	\end{align}
	where $\lambda$ is the trade-off parameter between the squared loss term and the regularizer representing the effect of the shape prior. \tcr{The term $\mathcal{L}^{\text{SP}}$ is carefully designed to simultaneously accomplish two tasks: 1.) a high detection rate of nuclei is encouraged since the element-wise Hadamard product highlights edge boundaries, and 2.) the subsequent convolution with the expert provided shape set reduces false positives.} Note that $\mathbf{\hat{y}}:=f(\mathbf{x}; \mathbf{\Theta})$, thus the shape prior cost term is effected by the network parameters and also introduces gradient terms that updates the network parameters during the training process using back-propagation \cite{lecun2015deep}. This is indicated in Fig. \ref{fig:diagram} by dashed line under ``Influencing CNN". Note that while the black, orange, and blue arrows are used during training, the black arrows are just used in testing stage.
	
	Let $\mathcal{L} = \mathcal{L}^{\text{Loss}} + \mathcal{L}^{\text{SP}}$ be the cost function. At iteration $t$ of the back-propagation, the filters of CNN are updated as: 
	
	\begin{equation}
	\label{eq:update}
	\mathbf{\Theta}^{t+1} = \mathbf{\Theta}^t - \eta \nabla_{\mathbf{\Theta}}\mathcal{L}
	\end{equation} 
	where $\eta$ denotes the learning rate, and $\mathbf{\Theta} = \{\mathbf{W}_k, \mathbf{b}_k\}_{k=1}^{D}$. The following gradients are to be computed\footnote{Note the update rules and the gradients for the bias terms are similar and are included with more details in the supplementary document \cite{webpage}.}:
	$$\frac{\partial \mathcal{L}}{\partial \mathbf{W}_k}, \frac{\partial \mathcal{L}}{\partial \mathbf{b}_k}$$
	where $\mathbf{W}_k$ and $\mathbf{b}_k$ denote one of the filters and biases at $k$-th layer of the CNN, representatively. The equation for computing the gradient w.r.t. an arbitrary entry within filter $\mathbf{W}_k$ in layer $k\in\{1,...,D\}$ is given by:
	\begin{align}
	\label{eq:pW}
	\frac{\partial \mathcal{L}}{\partial \mathbf{W}^a_k}& = -<(\hat{\mathbf{y}} - \mathbf{y}),\frac{\partial \mathbf{\hat{y}}}{\partial \mathbf{W}^a_k}>_F\\&\hspace{-5pt} - \lambda\sum_{i=1}^N<(g_{p}(\mathbf{\hat{y}})\odot \mathbf{\hat{x}}) \ast \mathcal{S}_i,g_p^{-1}\left(\frac{\partial \hat{\mathbf{y}}}{\partial \mathbf{W}^a_k}\right)\odot \mathbf{\hat{x}}*\mathcal{S}_i>_F\nonumber
	\end{align}
	where $\mathbf{W}^a_k$ denotes an arbitrary scalar entry within the representative filter $\mathbf{W}_k$, $<\cdot,\cdot>_F$ denotes the real value Frobenius inner product\footnote{For two real valued matrices $\mathbf{A}$ and $\mathbf{B}$ with same dimension, $<\mathbf{A},\mathbf{B}>_F := \sum_{i,j} A_{i,j}B_{i,j}$ where $i,j$ are the indexes of the entries.}, and $g_p^{-1}(\cdot)$ assigns the derivative to where activations come from, the ``winning unit", because other units in the previous layer's pooling blocks did not contribute to it. Hence derivative of all the other units is assigned to e zero. In Eq. (\ref{eq:pW}), $\frac{\partial \mathbf{y}}{\partial \mathbf{W}^a_k}$ is computed by following the standard backpropagation rule for each layer $k$ \cite{lecun1998gradient}.

	\section{Tunable Shape Priors with Convolutional Neural Networks (TSP-CNN)}
	\label{sec:TSPCNN}
	
\begin{figure*}[t!]
		\centering
		\includegraphics[width=0.94\linewidth]{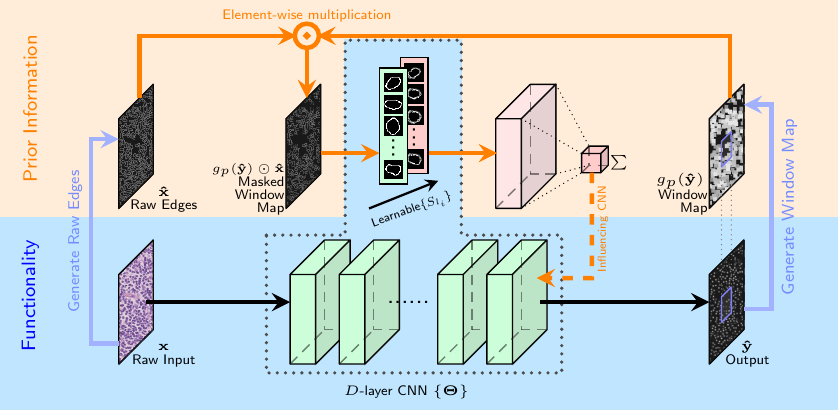}
		\caption{TSP-CNN illustration. There are two parts of TSP-CNN: Functionality part  and Prior Information part. Compared to SP-CNN, the shape set $\mathbf{S}_{l}$ is no longer fixed. It is now also a learnable component of the network. The shapes in the shape set are updated according to the shape learning regularization term as in Eq. (\ref{SSIM_cost}). As with SP-CNN, the prior information part generates regularization terms used in training the $D$ learnable layers of the CNN (through the blue and orange arrows) but {\em only} the functionality part of the network, i.e. the $D$ layer CNN (excluding the learned shapes) is applied to a test image to predict nuclei centers (through the black arrows).\vspace{-5pt}} 
		\label{fig:diagramNew}
	\end{figure*}

	While exploiting a shape set provided by a domain expert is meritorious, a fundamental open question is: What is the best, most compact shape set to employ and can it be adapted based on the image characteristics? This section provides an answer to that question in the following manner: the set of shapes can be seen as a collection of filters, which we now add to the network as a learnable component and call it the learnable shape layer -- see Fig. \ref{fig:diagramNew} for an illustration of this idea.  We call this new network structure as Tunable Shape Priors with CNNs (TSP-CNN). 
	
\begin{figure}[t]
		\centering
		\includegraphics[width=0.8\linewidth]{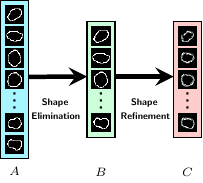}
		\caption{An illustration of the two-stage process of shape elimination and refinement. Starting with the domain expert provided shape set, a redundancy removing elimination step is performed first using Algorithm \ref{alg:elimination}. Second, refinement of this reduced size shape set is performed which adapts to training imagery and arrives at a compact {\em Basis Shape Set}.}
		\label{fig:diagramShapes}
	\end{figure}

	We observe that the domain expert provided shape set in practice is quite redundant, i.e.\ it may contain pairs of hand-crafted shapes that are quite close to each other and do not add value to nucleus detection. We therefore propose a two stage process to arrive at a compact {\em Basis Shape Set} which is adapted based on training imagery: 1.) shape elimination stage to obtain a reduced/smaller reference shape set, and 2.) shape refinement or learning via TSP-CNN. This two stage process is illustrated in Fig. \ref{fig:diagramShapes}. For both stages, meaningful shape similarity measures are needed; the exact shape elimination and refinement procedure is detailed next. We next introduce shape elimination and refinement procedures, and then describe the training procedure to learn shapes that are adapted to the underlying dataset.
	
	\subsection{Shape Elimination}
	\label{sec:shapeElimination}
	
	Many shape similarity measures have been developed in image processing and vision for a variety of different tasks \cite{veltkamp2001shape,sampat2009complex,attalla2005robust,latecki2000shape,lu1999region,super2004learning,ueda1993learning}. Most of these methods measure the distance between shapes $A$ and $B$, $d(A,B)$ in different ways, for example bottleneck distance, Hausdroff distance, turning function distance, etc \cite{veltkamp2001shape}. Attalla \textit{et al.} used a multi-resolutional polygonal shape descriptor that is invariant to scale and is used to match and recognize 2D shapes. In \cite{latecki2000shape}, Latecki \textit{et al.} proposed a cognitively motivated similarity measure which divides a shape into best possible correspondences of visual parts and measures the similarity between each of them and then aggregates them to find the overall similarity measure.
	
	In \cite{sampat2009complex}, Sampat \textit{et al.} proposed Complex Wavelet SSIM (CW-SSIM) which is an extension of the well-known Structural Similarity Index Measure (SSIM) \cite{Wang-SSIMTut}. CW-SSIM has been shown to be insensitive to nonstructural geometric distortions and an excellent measure to compare binary shape similarity \cite{sampat2009complex} because of its ability to capture phase information. CW-SSIM index \cite{sampat2009complex} for two sets of coefficients in the complex wavelet transform domain as $\mathbf{c}_x = \{c_{x,i}|i=1,\dots, M\}$ and $\mathbf{c}_y = \{c_{y,i}|i=1,\dots, M\}$, extracted from same location of same wavelet subbands of two shape images being compared is defined as:
	\begin{equation}
	\textrm{CW-SSIM}(\mathbf{c}_x, \mathbf{c}_y) = \frac{2|\sum_{i=1}^{M}c_{x,i}c^*_{y,i}| + K}{\sum_{i=1}^{M}|c_{x,i}|^2 + \sum_{i=1}^{M}|c_{y,i}|^2 + K},
	\label{eq:cwssim}
	\end{equation}
	where $c^*$ is the complex conjugate of $c$ and $K$ is a small positive constant. After computing CW-SSIM values for all sets, $\mathbf{c}_i$'s, the final CW-SSIM value for two shape images $\textrm{CW-SSIM}(\mathcal{S}_i,\mathcal{S}_{j})$ is obtained by averaging CW-SSIM value over all wavelet subbands as described in \cite{sampat2009complex}.

	\begin{algorithm}[t!]
		\caption{Shape elimination procedure using CW-SSIM}
		\label{alg:elimination}
		\begin{algorithmic}[1]
			\renewcommand{\algorithmicrequire}{\textbf{Input:}}
			\renewcommand{\algorithmicensure}{\textbf{Output:}}
			\REQUIRE $\mathbf{S} = \{\mathcal{S}_i | i = 1, 2, \dots, N\}$, $k, n=1$, $\widehat{\mathbf{S}}, \mathbf{S}_n = \{\}$\footnotemark 
			%	\\ \textit{Initialisation} :
			%	\STATE first statement
			%	\\ \textit{LOOP Process}
			\WHILE {$k = 1$}
			\STATE $\mathbf{S}_n \gets \mathbf{S}_n \cup \{\mathcal{S}_k$\}
			\FOR {$l = 2$ to $N$}
			\STATE $\mathcal{C} = \text{CW-SSIM}(\mathcal{S}_k,\mathcal{S}_l)$
			\IF {$\mathcal{C}>0.8$}
			\STATE $\mathbf{S}_n \gets \mathbf{S}_n\cup\{\mathcal{S}_l\}$
			\ELSE
			\STATE Go to line 3
			\ENDIF
			\ENDFOR
			\STATE $\mathbf{S} \gets \mathbf{S}\setminus \{\mathbf{S}\cap\mathbf{S}_n\}$
			\STATE $\widehat{\mathbf{S}} \gets \widehat{\mathbf{S}} \cup \{\text{Rep}(\mathbf{S}_n)\}$
			\STATE $n \gets n+1$
			\STATE $N = |\mathbf{S}|$
			\IF {$N$ = 0}
			\STATE $k=0$
			\ENDIF
			\ENDWHILE
			%	\RETURN $P$ 
			\ENSURE  Reference shape set $\mathbf{S}_r = \widehat{\mathbf{S}}$.
		\end{algorithmic} 
	\end{algorithm}
	\footnotetext{In Algorithm \ref{alg:elimination}, $\{\}$ is an empty set, $|\mathbf{S}|$ is the cardinality of set $\mathbf{S}$, and $\text{Rep}(\mathbf{S}_n)$ is one representative shape from grouped similar shapes in set $\mathbf{S}_n$.}
	
	Starting with the domain expert provided shape set $\mathbf{S} = \{\mathcal{S}_{i} | i = 1, 2, \dots, N\}$, we eliminate redundancy by using the aforementioned CW-SSIM measure to compute pairwise comparisons of shapes. Near identical (redundant) shapes are grouped together and one representative is extracted from the group. The complete procedure is detailed in Algorithm  \ref{alg:elimination}.\vspace{-5pt}
	
	\subsection{Shape Refinement}
	\label{sec:shapeRefinement}
	
	The output of Algorithm  \ref{alg:elimination} is what we address as $\mathbf{S}_r = \{\mathcal{S}_{r_j} | j = 1, 2, \dots, Q\}$ the Reference Shape Set -- which is essentially a cardinality (size) reduced version of the domain expert provided shape set $\mathbf S$, i.e.\ $Q < N$. Our approach to learn or refine shapes is then to optimize a Learnable Shape Set $\mathbf{S}_l = \{\mathcal{S}_{l_i} | i = 1, 2, \dots, Q\}$ under the important physical constraint that the shapes in $\mathbf{S}_l$ bear similarity to those in $\mathbf{S}_r$. 
	\begin{figure}[t]
		\centering
		\includegraphics[width=0.45\textwidth]{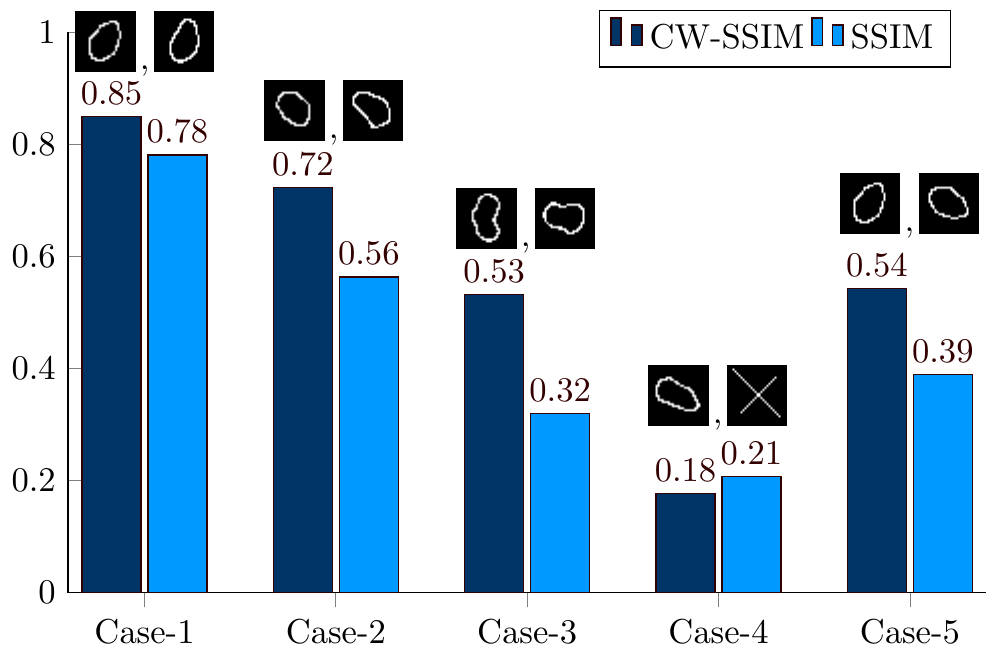}
		\caption{SSIM \& CW-SSIM values for different cases of shapes}
		\label{fig:ssim_chart}
	\end{figure}
	
	Ideally, as in the shape elimination stage, CW-SSIM should be used as a similarity measure between $\mathbf{S}_l$ and $\mathbf{S}_r$. However, CW-SSIM is not differentiable and  hence arduous to incorporate in standard CNN learning frameworks which rely on derivative based back-propagation. SSIM, a pre-cursor to CW-SSIM on the other hand is differentiable and implementable in a deep learning framework \cite{zhao2017loss}. In Fig. \ref{fig:ssim_chart}, we compare CW-SSIM and SSIM values in four cases of binary shape similarity by using representative nuclei shapes corresponding to the image datasets we work with: case-1: very similar, case-2: similar, case-3: different, case-4: very different, \tcr{and case-5: rotation ($90^{\circ}$)}.  From Fig. \ref{fig:ssim_chart} we can infer two facts: 1.) CW-SSIM can be quite effective in comparing binary nuclei shapes, and 2) SSIM forms an approximation to CW-SSIM. Given the tractability benefits of SSIM in deep learning set-ups, we employ it to build our shape similarity regularizer. As is corroborated later in Section \ref{sec:experiments}, using an SSIM based regularizer suffices to afford TSP-CNN significant practical gains. \tcr{The design of more sophisticated e.g. rotation invariant, similarity measures that can be tractably incorporated in deep learning frameworks is a viable direction for future research}.

	The SSIM index between patches $\mathbf{x}\in\mathcal{S}_{l_i}$ and $\mathbf{y}\in\mathcal{S}_{r_j}$ is
	\begin{equation}
	\textrm{SSIM}(\mathbf{x},\mathbf{y}) = \frac{2\mu_{\mathbf{x}}\mu_{\mathbf{y}}+C_1}{\mu^2_{\mathbf{x}} + \mu^2_{\mathbf{y}} + C_1}.\frac{2\sigma_{\mathbf{x}\mathbf{y}}+C_2}{\sigma^2_{\mathbf{x}} + \sigma^2_{\mathbf{y}} + C_2},
	\label{eq:ssim}
	\end{equation}
	where $\mu$ and $\sigma$ are mean and standard deviation of the patches, respectively. The patch selection, comparisons and estimation of local statistics is done as in \cite{Wang-SSIMTut}. The overall SSIM value between two shapes, $\textrm{SSIM}(\mathcal{S}_{l_i}, \mathcal{S}_{r_j})$, is the average of SSIM value over all the patches in the shapes. 
	
	In \cite{zhao2017loss}, the use of SSIM is motivated by image quality concerns. In TSP-CNN, differently from \cite{zhao2017loss}, we use SSIM to define a shape learning regularizer in the following manner:
	\begin{equation}
	\mathcal{L}^{\textrm{SSIM}} = -\gamma\textrm{SSIM}(\mathbf{S}_l,\mathbf{S}_r) = -\gamma \sum_{i=1}^{Q}\sum_{j=1}^{Q}\textrm{SSIM}(\mathcal{S}_{l_i}, \mathcal{S}_{r_j}) \label{SSIM_cost}
	\end{equation}
	and optimize the network parameters (including shapes) with:
	\begin{align}
	\widehat{\mathbf{\Theta}} = &\arg\min\limits_{\widehat{\mathbf{\Theta}}} \mathcal{L}^{\text{Loss}} + \mathcal{L}^{\text{SP}} + \mathcal{L}^{\text{SSIM}} \nonumber\\
	= &\arg\min\limits_{\widehat{\mathbf{\Theta}}} \|f(\mathbf{x}; \widehat{\mathbf{\Theta}}) - \mathbf{y}\|_2^2 - \lambda\sum\limits_{i=1}^{Q}   \|(g_{p}(\mathbf{\hat{y}})\odot \mathbf{\hat{x}}) \ast \mathcal{S}_{l_i}  \|_2^2 \nonumber\\
	& - \gamma \sum_{i=1}^{Q}\sum_{j=1}^{Q}\textrm{SSIM}(\mathcal{S}_{l_i}, \mathcal{S}_{r_j})
	\label{finalCostFunc}
	\end{align}
	where $\widehat{\mathbf{\Theta}} = \{\mathbf{\Theta}, \mathbf{S}_l\} = \{\mathbf{W}, \mathbf{b}, \mathbf{S}_l\}$. In Eq. \ref{finalCostFunc}, $\mathcal{L}^{\text{Loss}}$ is the standard mean squared loss term between the ground-truth and output of the network. 
	
	Note that reference shape set $\mathbf{S}_r$ is obtained using the Algorithm \ref{alg:elimination}, that is $\mathbf{S}_r$ is a pruned shape set obtained using the shape elimination procedure. The shapes that are actually learned are in the set $\mathbf{S}_l$, and therefore $\mathbf{S}_l$ once optimized represents the refined version of the shapes in $\mathbf{S}_r$.

	The shapes in set $\mathbf{S}_l$ are learned according to the shape prior term ($\mathcal{L}^{\text{SP}}$) and the shape similarity measure ($\mathcal{L}^{\text{SSIM}}$): in particular, as argued before the shape prior term $\mathcal{L}^{\text{SP}}$ should be minimized -- so in essence, we are looking for the best shape set $\mathbf{S}_l$ that can do so. Note that the second regularizer $\mathcal{L}^{\text{SSIM}}$ emphasizes similarity to the pruned reference shape set $\mathbf{S}_r$ and therefore ensures that physically meaningful shapes are learned -- unconstrained optimization of $\mathbf{S}_l$ could lead to learned shapes that do not conform to reality.
	
	\subsection{Back-propagation: Shape Learning Cost Term}
	The back-propagation for TSP-CNN is similar to that of SP-CNN, except for the new shape learning term. The back-propagation for shape learning term only depends on the shapes $\mathcal{S}_{l_i}$. Let the cost function be $\widehat{\mathcal{L}} = \mathcal{L}^{\text{Loss}} + \mathcal{L}^{\text{SP}} + \mathcal{L}^{\text{SSIM}}$, hence its gradient w.r.t. $\mathcal{S}^a_{l_i}$, an arbitrary scalar entry within the shape $\mathcal{S}_{l_i}$, is as follows:
	\begin{align}
	\frac{\partial\widehat{\mathcal{L}}}{\partial \mathcal{S}_{l_i}^a} = &-\lambda\sum_{i=1}^Q <(g_{p}(\mathbf{\hat{y}})\odot \mathbf{\hat{x}}) \ast \mathcal{S}_{l_i},(g_{p}(\mathbf{\hat{y}})\odot \mathbf{\hat{x}}) \ast \mathbf{I}>_F \nonumber\\& -\gamma\sum_{j=1}^{Q}\frac{\partial}{\partial \mathcal{S}_{l_i}^a}\textrm{SSIM}(\mathcal{S}_{l_i},\mathcal{S}_{r_j})\label{eq:bkp}
	\end{align}
	where $\mathbf{I}$ denotes a matrix with same size as $\mathcal{S}_{l_i}$ with only entrance $a$ active as one. Second term in right hand side of Eq. \ref{eq:bkp} is derived in detail in \cite{wang2008maximum,zhao2017loss}. For a complete derivation of the back-propagation procedure, refer to our supplementary document in \cite{webpage}.

	\begin{figure}[t]
		\centering
		\includegraphics[width=0.5\textwidth]{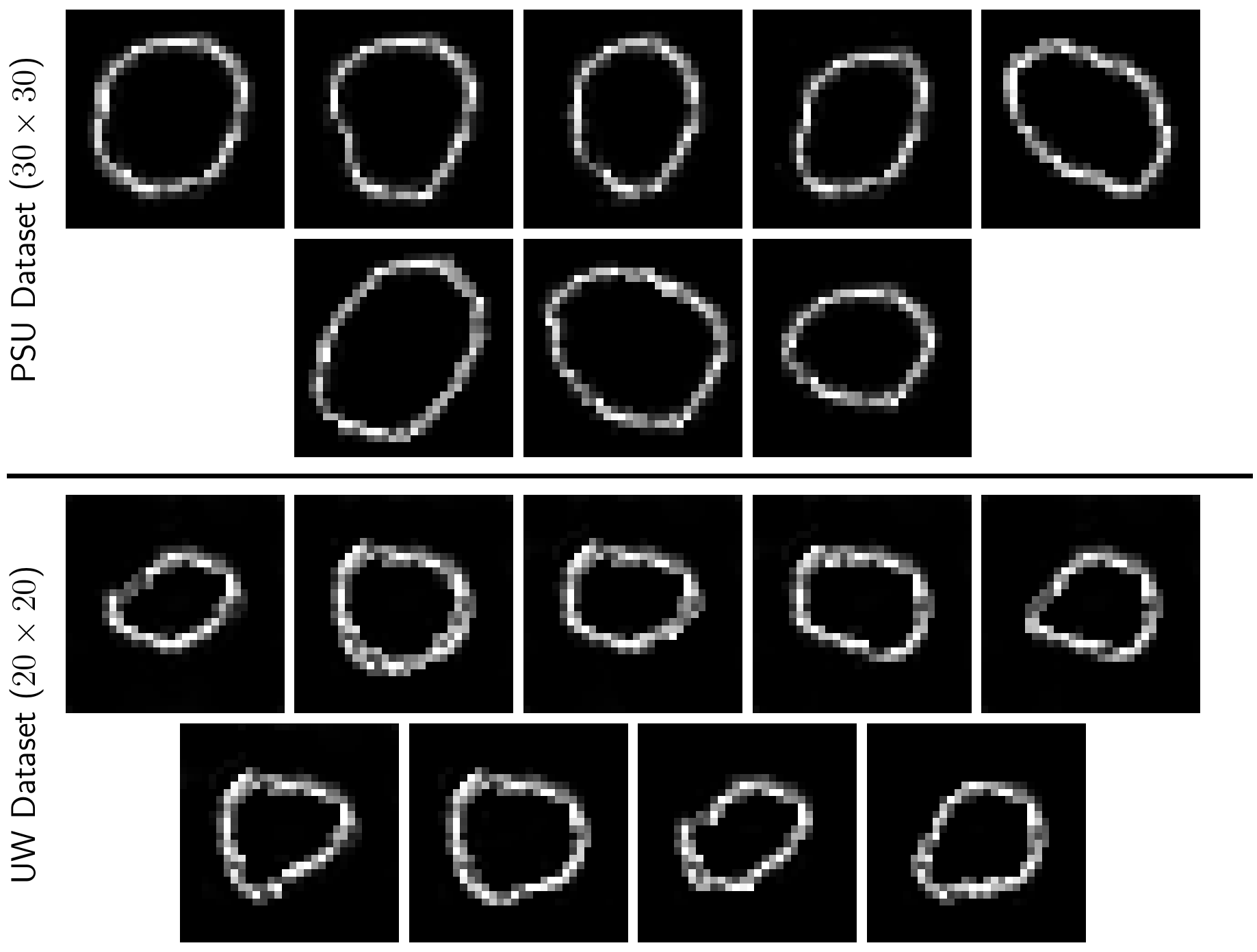}
		\caption{Learned basis shapes using TSP-CNN for both UW and PSU datasets. Note that our optimization of shapes does not constrain them to be binary and hence these shapes have thicker boundaries with values in between 0 and 1.}
		\label{fig:outputShapes}
	\end{figure}
	
	\section{Experimental Results}
	\label{sec:experiments}
	\subsection{Experimental Setup and Datasets}
	\label{sec:data}
\begin{figure}[t!]
		\subfloat[PSU Dataset]{\includegraphics[width=.27\textwidth]{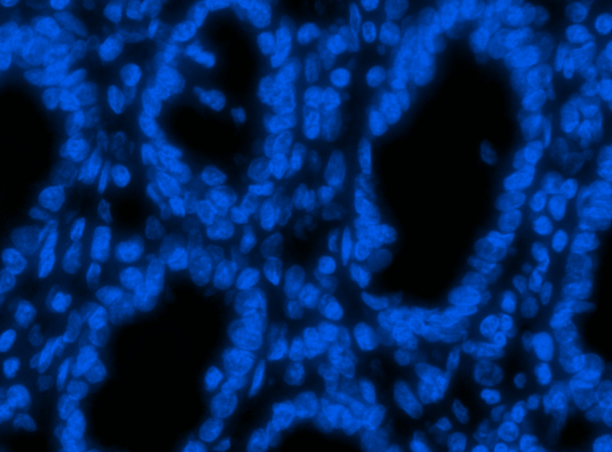}}\hspace{.1mm}
		\subfloat[UW Dataset \cite{SC_CNN}]{\includegraphics[width=.2\textwidth]{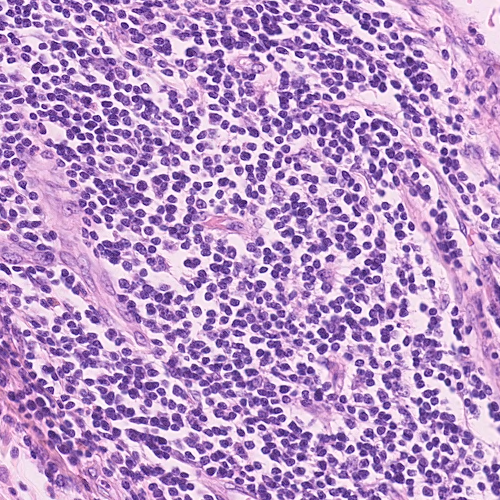}}\hspace{.1mm}
		\caption{Sample images from the datasets used for evaluation}
		\label{fig:sampleImages}
	\end{figure}

	We train and test SP-CNN and TSP-CNN on two colon cancer datasets: 1) publicly available UW Dataset \cite{SC_CNN} which includes $100$ H\&E stained histology images of colorectal adenocarcinomas. There are a total number of $29756$ nuclei marked at the nucleus center (please refer to Sec. VII.A. of \cite{SC_CNN} for more information). Our choice of the UW Dataset is because it represents real-world challenges such as overlapping nuclei and contains other shapes that are often confused with nuclei. Further, it is one of the few publicly available datasets that is widely used in many recent deep learning-based nuclei detection methods \cite{SC_CNN, CSP_CNN, Ahmad2017}, 2) a new dataset carefully prepared and labeled manually by medical experts at the Center for Molecular Immunology and Infectious Disease, Penn State University. We call it the `PSU Dataset' and it includes $120$ images of colon tissue from $12$ pigs at a resolution of $0.55 \mu$m/pixel. Formalin fixed paraffin embedded pig colon sections were deparaffinized and stained with fluorescent DNA stain DAPI (4',6-diamidino-2-phenylindole) to visualize the cells as described in \cite{Charepalli2017}. The selected images represent cross-sectional view of the colon epithelial cells. It also comprised of areas with artifacts, over-staining, and failed auto focusing, to represent outliers normally found in real scenarios. A total number of $25462$ nuclei are annotated manually by an expert. For reproducing research results, we have made this dataset publicly available at the SP-CNN web-page \cite{webpage}. Sample images from both datasets are shown in Fig. \ref{fig:sampleImages}.

\begin{figure}[t]
		\centering
		\hspace{-2pt}\includegraphics[width=0.48\textwidth]{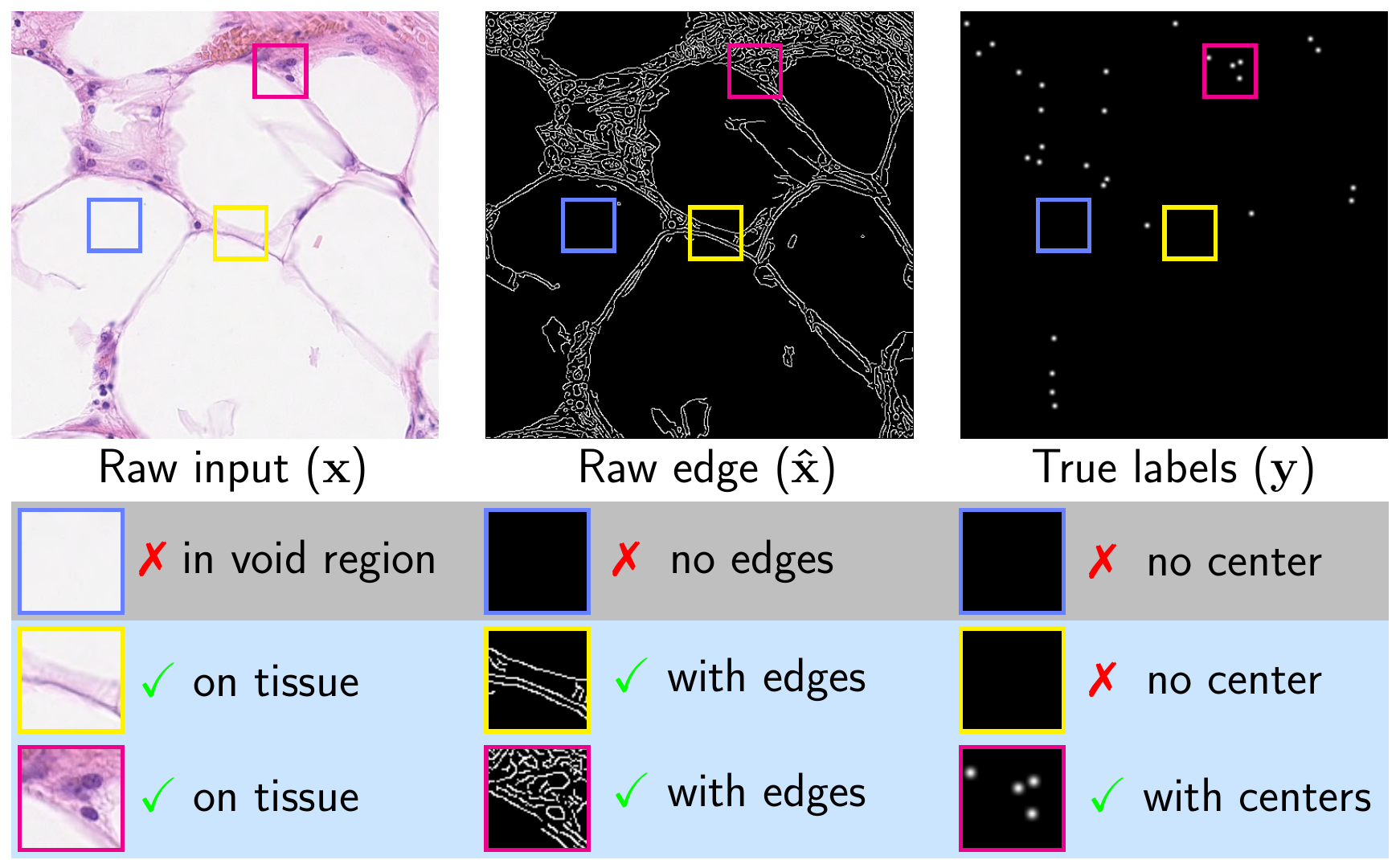}
		\caption{Void training tuple (raw image $\mathbf{x}$, raw edge image $\mathbf{\hat{x}}$, ground-truth label image $\mathbf{y}$) elimination procedure.}
		\label{fig:sampleSelection}
	\end{figure}

	We construct $\mathbf{y} \in [0,1]$ by processing the binary image of ground-truth nuclei center locations, which has 1 at the nucleus center and 0 elsewhere. This is accomplished by convolving the said ground-truth binary image with a zero mean Gaussian ($\sigma = 2$) filter of size $7 \times 7$. Then the (luminance) input image, the raw edge image, and the labeled image form a training tuple $(\mathbf{x}, \mathbf{\hat{x}}, \mathbf{y})$; patches of size $40\times 40$ are extracted and used for training. There are void regions in the raw image $\mathbf{\hat{x}}$, which do not include nuclei, as a result the label image $\mathbf{y}$ is also void. These redundant void regions can mislead the network while wasting computation. To avoid this, we develop a procedure to eliminate the training patches which are empty in $(\mathbf{x}, \mathbf{\hat{x}}, \mathbf{y})$ -- this is visually illustrated in Fig. \ref{fig:sampleSelection}.

	\subsection{Assessment Metrics and Parameters}
	\label{sec:assessment}
	The output of SP/TSP-CNN is normalized as $\mathbf{\hat{y}} \in [0,1]$, which is then processed via a thresholding operation with a pre-determined threshold $T$. Local maxima of the resulting thresholded image are identified as detected nuclei locations. To evaluate the detected locations against the true ones, we need some tolerance since it is unlikely that they will exactly match. This is handled in the literature by defining a golden standard region around each ground-truth nuclei center as described in \cite{SC_CNN}: we define this to be a region of 6 pixels around each nuclei center for UW Dataset and a region of 10 pixels for PSU Dataset (since the nuclei are larger in this dataset). Note that for fairness,  the {\em same} `Golden Region' is used across all methods that are compared in Section \ref{sec:expD}.
	
	A detected nuclei location is considered to be true positive ($TP$), if it lies inside this region, otherwise it is considered to be false positive ($FP$), and the ones that are not matched by any of golden standard regions are considered to be false negative ($FN$). For quantitative assessment of SP-CNN and TSP-CNN and comparison with other methods we use Precision ($P$), Recall ($R$), and F1 score ($F1$) which are
	\label{sec:params}
	\begin{equation}
	\small P = \frac{TP}{TP + FP}, ~R = \frac{TP}{TP + FN}, ~\text{and}~ F1 = \frac{2PR}{P+R}.
	\end{equation}
	To determine parameters for both networks, we use cross validation \cite{monga2017}, which leads to the following choice\footnote{During the training, both SP- and TSP-CNN network parameters are initialized using the same random seed to make for a fair comparison.}:
	
	\noindent \textbf{SP-CNN:} $D=7$ layers are used as validated in Section \ref{layers}. These layers are with `SAME' padding scheme; its configuration details are provided in Table. \ref{tab:network}. We use $N = 24$ different nuclei shapes in the domain expert provided shape set. Each shape is described by a $20 \times 20$ patch for UW Dataset \cite{SC_CNN}, while for the PSU dataset, we use a $30 \times 30$ patch. The active part of the shapes are labeled as 1 and 0 otherwise. All the parameters used in SP-CNN are chosen by cross validation \cite{monga2017}. Most important of them are: trade-off values $\lambda~=~5e-7$ and $\gamma = 0$, pooling window size $p = 11\times 11$, weight decay parameter = $1e-5$, learning rate decay = $0.75$, where the trade-off parameters are selected by cross-validation and the effect of these parameters is shown in Table \ref{tab:parameters}. 
	
	\noindent \textbf{TSP-CNN:}	The parameters in TSP-CNN are same as SP-CNN except for the trade-off parameters which are $\lambda~=~1e-10$ or $1e-12$ and $\gamma = 2$, and number of layers which is chosen (as per Section \ref{layers}) to be $D=10$. With this parameter setting, the Basis Shape Set using TSP-CNN for the UW-dataset converges to $Q = 9$ basis shapes, while for the PSU dataset $Q = 8$ shapes are determined. For both datasets, the optimized {\em Basis Shape Set} is visualized in Fig.\ \ref{fig:outputShapes}.

	\begin{table}[t]
		\centering
		\caption{Effect of different parameters (PSU Dataset)}
		\label{tab:parameters}
		\resizebox{1\linewidth}{!}{
			\begin{tabular}{cc|cc||cc|cc}
				\hline\hline
				\multicolumn{4}{c||}{SP-CNN}     & \multicolumn{4}{c}{TSP-CNN (p=11x11)} \\ \hline
				$\gamma$ & F1 & p & F1 & $\lambda$ & F1 & $\gamma$ & F1 \\ \hline
				1e-6 & 0.777 & 5x5 & 0.806 & 0.01 & 0.803 & 1e-6 & 0.824 \\ 
				\textbf{5e-7} & \textbf{0.863} & 7x7 & 0.762 & 0.1 & 0.807 & 1e-8 & 0.818 \\ 
				1e-8 & 0.818 & 9x9 & 0.809 & \textbf{2} & \textbf{0.892} & \textbf{1e-10} & \textbf{0.892} \\ 
				1e-12 & 0.731 & \textbf{11x11} & \textbf{0.863} & 10 & 0.804 & 1e-12 & 0.812 \\ \hline\hline
		\end{tabular}}
	\end{table}
	
	\begin{table}[t]
		\centering
		\caption{The Configuration of The CNN Used in SP-CNN}
		\label{tab:network}
		\resizebox{1\linewidth}{!}{
			\begin{tabular}{cccc}
				\hline\hline
				Layer No. & Layer Type & Filter Dimensions & Filter Numbers\\ \hline
				1 & Conv. + ReLU & $5\times 5\times 1$ &$64$ \\ 
				2 & Conv. + ReLU & $3\times 3\times 64$ &$64$ \\ 
				3 & Conv. + ReLU & $3\times 3\times 64$ &$64$ \\ 
				4 & Conv. + ReLU & $3\times 3\times 64$ &$64$ \\ 
				5 & Conv. + ReLU & $3\times 3\times 64$  &$64$\\ 
				6 & Conv. + ReLU & $3\times 3\times 64$ &$64$ \\ 
				7 & Conv. + ReLU & $3\times 3\times 64$ &$64$ \\ 
				8 & Conv. + ReLU & $3\times 3\times 64$ &$64$ \\ 
				9 & Conv. + ReLU & $3\times 3\times 64$  &$64$\\ 
				10 & Convolutional & $3\times 3\times 64$  &$1$\\ \hline\hline
		\end{tabular}}
	\end{table}

	\subsection{Determining Number of Layers for the Network}
	\label{layers}
\begin{figure}[t]
		\centering
		\includegraphics[width=0.47\textwidth]{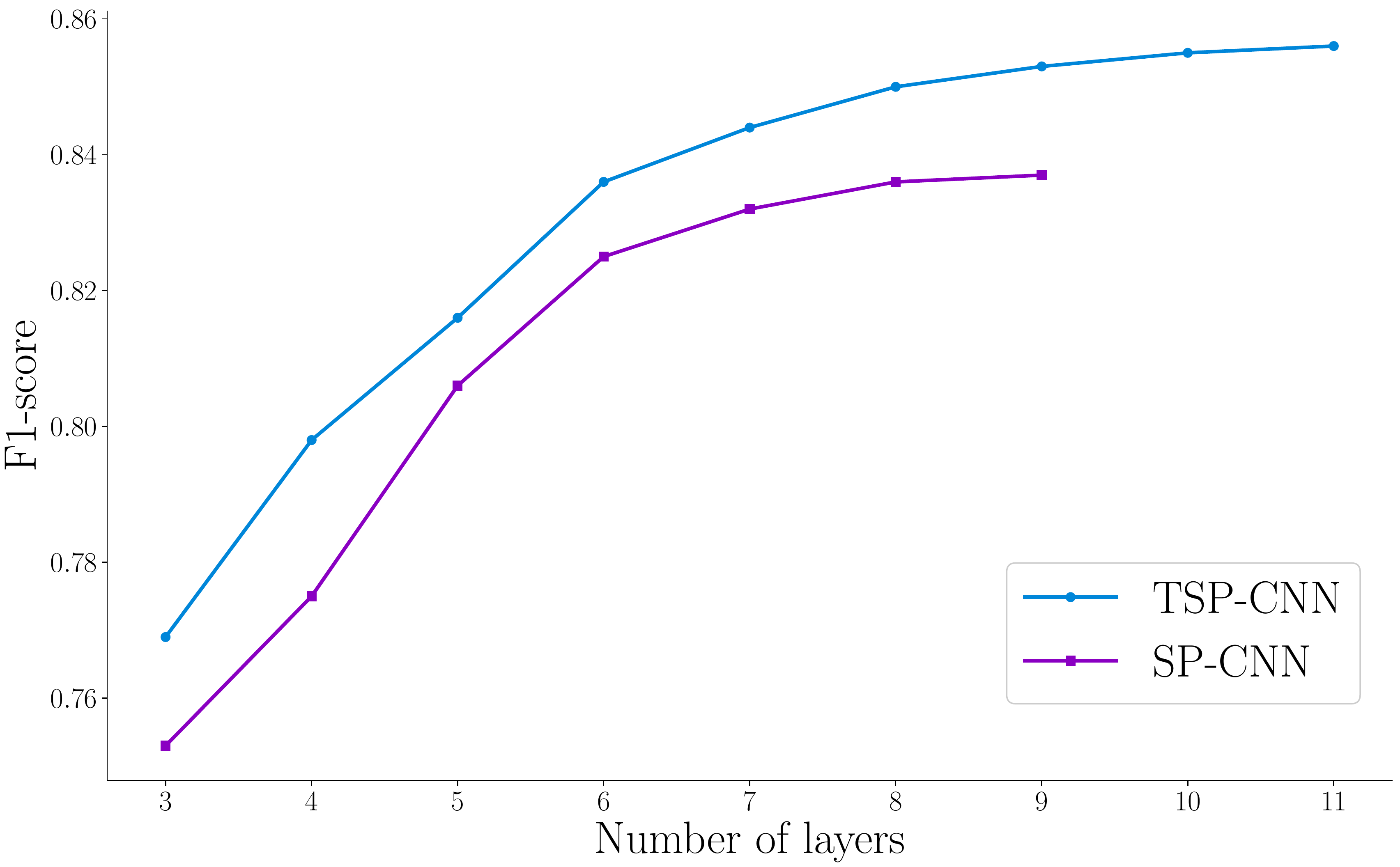}
		\caption{F1-score for different number of layers: UW Dataset.}
		\label{fig:D_layer}
	\end{figure}

	In order to come up with the number of layers which suffices for accurate nuclei detection, we train the network with layers $D=3$ to $D=11$. The impact of number of layers is investigated in Fig. \ref{fig:D_layer} for both SP-CNN and TSP-CNN. Fig. \ref{fig:D_layer} reveals that $D=10$ for TSP-CNN and $D=7$ layers for SP-CNN facilitate the best computation-performance balance, i.e. more layers can mildly help but also increase computational burden. Other competing deep networks also employ 6-10 layers \cite{SC_CNN,Xing2016}.
	
	TSP-CNN produces better results than SP-CNN even using shallower CNNs because the learned shapes adapt to the dataset. Note from Fig. \ref{fig:D_layer} that a 6 layer TSP-CNN has the same level of performance as the 9 layer SP-CNN. Finally, while Fig. \ref{fig:D_layer} is plotted for the UW Dataset, we found similar trends for the PSU dataset.

	\begin{figure}
		\centering
		\includegraphics[width=0.49\textwidth]{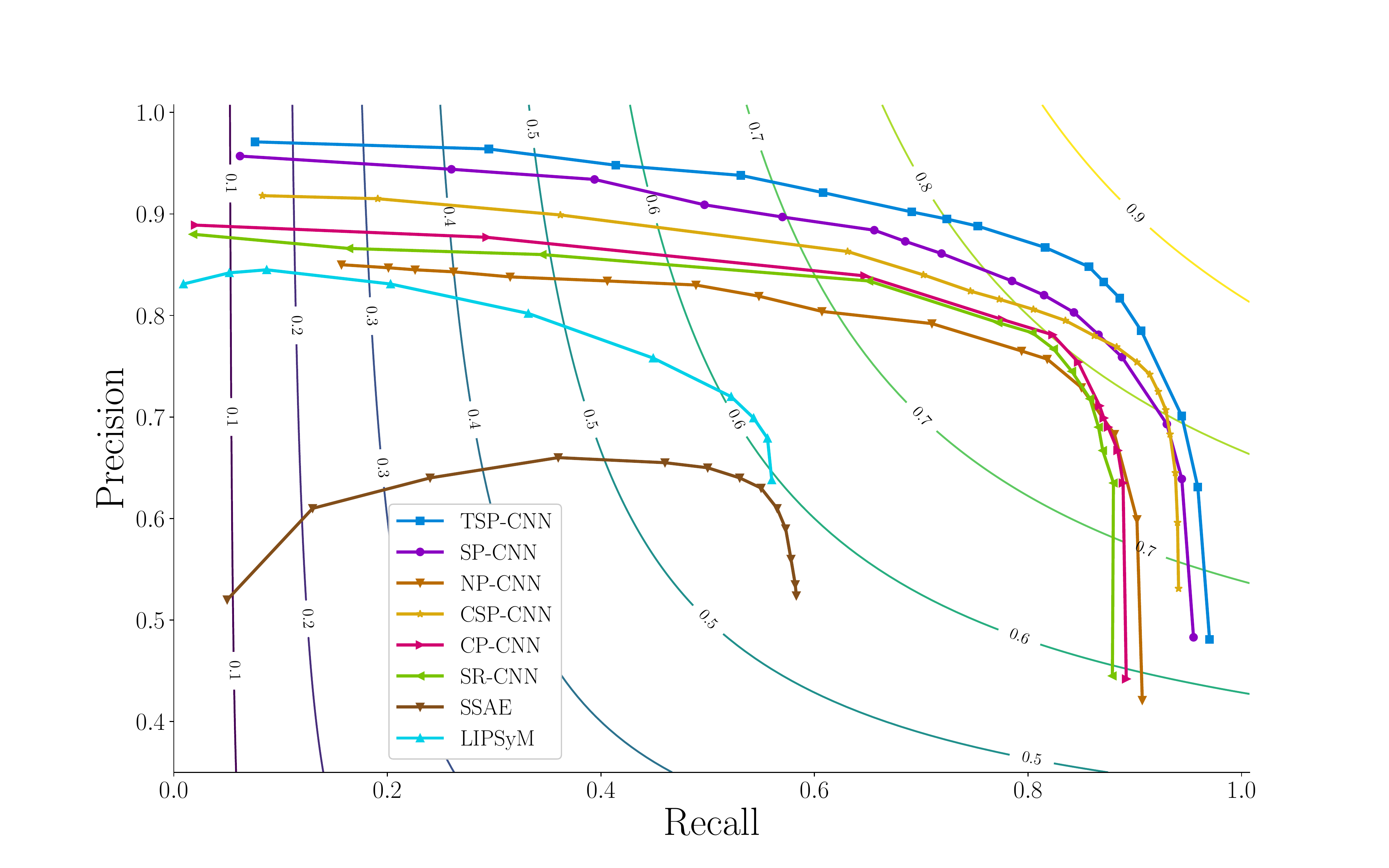}
		\caption{Precision-recall curve for UW Dataset \cite{SC_CNN}.}
		\label{fig:PRcurve}
	\end{figure}%
	
	\begin{figure}
		\centering
		\includegraphics[width=0.49\textwidth]{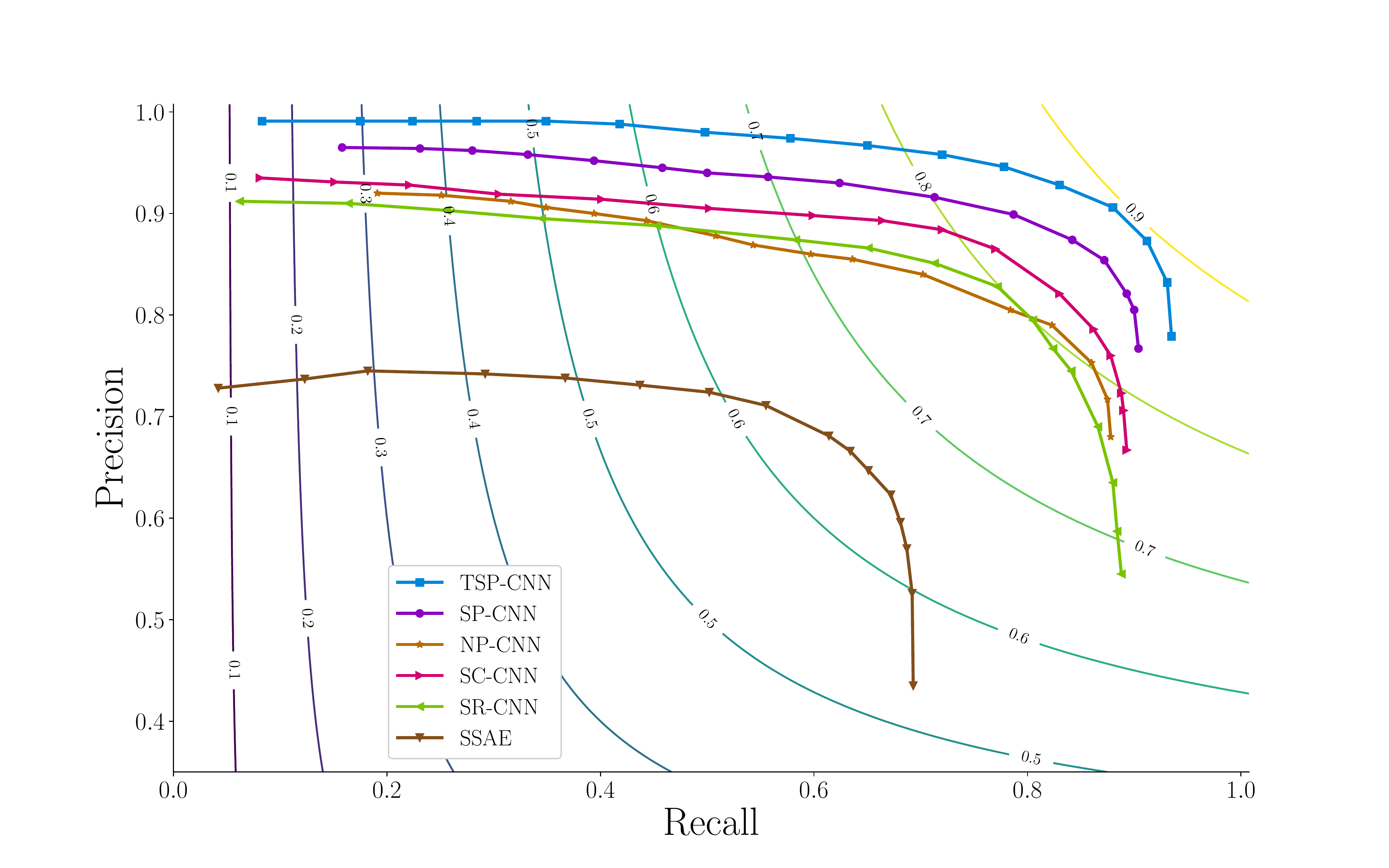}
		\caption{Precision-recall curve for PSU Dataset.}
		\label{fig:PRcurvePSU}
	\end{figure}%

	\begin{table}
		\centering
		\caption{Nucleus detection results for UW Dataset \cite{SC_CNN}}
		\label{tab:results}
		\resizebox{0.8\linewidth}{!}{
			\begin{tabular}{cccc}
				\hline	\hline
				UW Dataset \cite{SC_CNN} & Precision & Recall & F1 score \\ \hline
				TSP-CNN & 0.848 & 0.857 & \textbf{0.852} \\
				SP-CNN & 0.803 & 0.843 & 0.823 \\ 
				\tcr{WNP-CNN} & 0.761 & 0.829 & 0.793 \\
				NP-CNN & 0.757 & 0.818 & 0.786 \\  
				CSP-CNN \cite{CSP_CNN} & 0.788 & 0.886 & 0.834 \\ 
				SC-CNN \cite{SC_CNN} & 0.781 & 0.823 & 0.802 \\
				SR-CNN \cite{Xie2015} & 0.783 & 0.804 & 0.793 \\
				SSAE \cite{SSAE} & 0.617 & 0.644 & 0.630 \\
				LIPSyM \cite{LIPSyM} & 0.725 & 0.517 & 0.604 \\ 
				CRImage \cite{CRImage} & 0.657 & 0.461 & 0.542\\ \hline\hline
		\end{tabular}}
	\end{table}%
	
	\begin{table}
		\centering
		\caption{Nucleus detection results for PSU dataset}
		\label{tab:results1}
		\resizebox{0.8\linewidth}{!}{
			\begin{tabular}{cccc}
				\hline	\hline
				PSU Dataset & Precision & Recall & F1 score \\ \hline
				TSP-CNN & 0.874 & 0.911 & \textbf{0.892} \\ 
				SP-CNN & 0.854 & 0.871 & 0.863 \\ 
				NP-CNN & 0.746 & 0.859 & 0.799 \\ 
				SC-CNN \cite{SC_CNN} & 0.821 & 0.830 & 0.825 \\ 
				SR-CNN\cite{Xie2015} & 0.797 & 0.805 & 0.801 \\
				SSAE\cite{SSAE} & 0.665 & 0.634 & 0.649 \\ \hline\hline
		\end{tabular}}
	\end{table}%
	
	\begin{figure}
		\centering
		\includegraphics[width=0.487\textwidth]{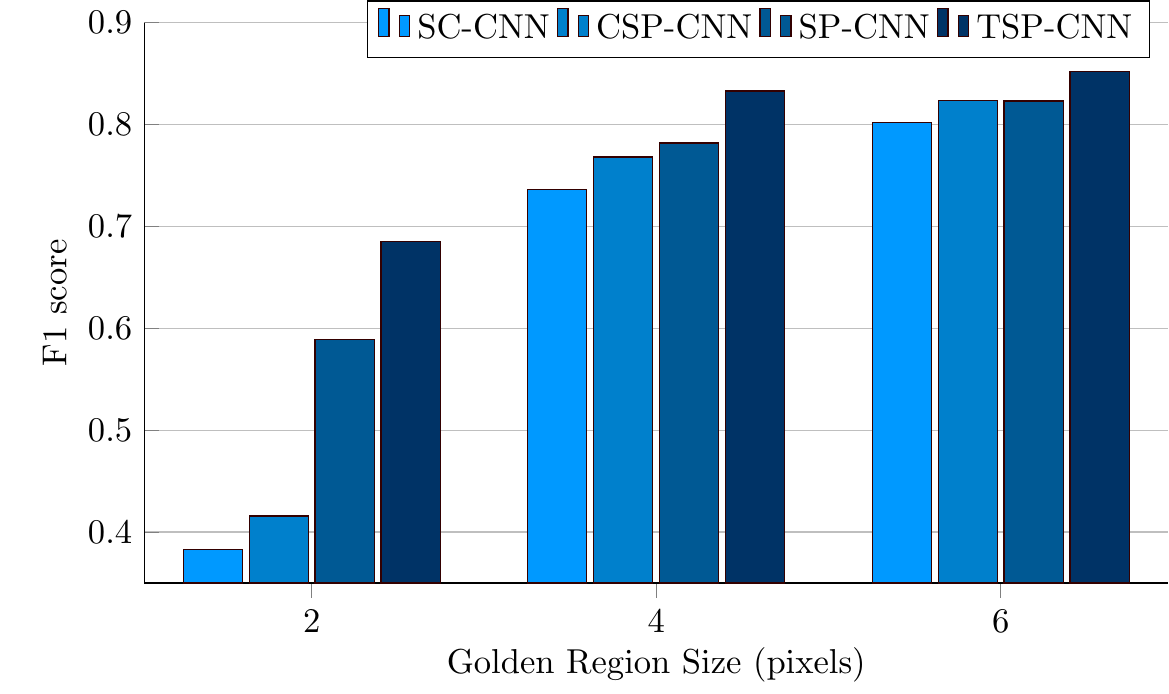}
		\caption{F1 scores for the top competing methods over varying choice of the Golden Region for the UW Dataset \cite{SC_CNN}.}
		\label{fig:GR}
	\end{figure}%

	\begin{figure*}[ht!]
		\centering
		\includegraphics[width=1\textwidth]{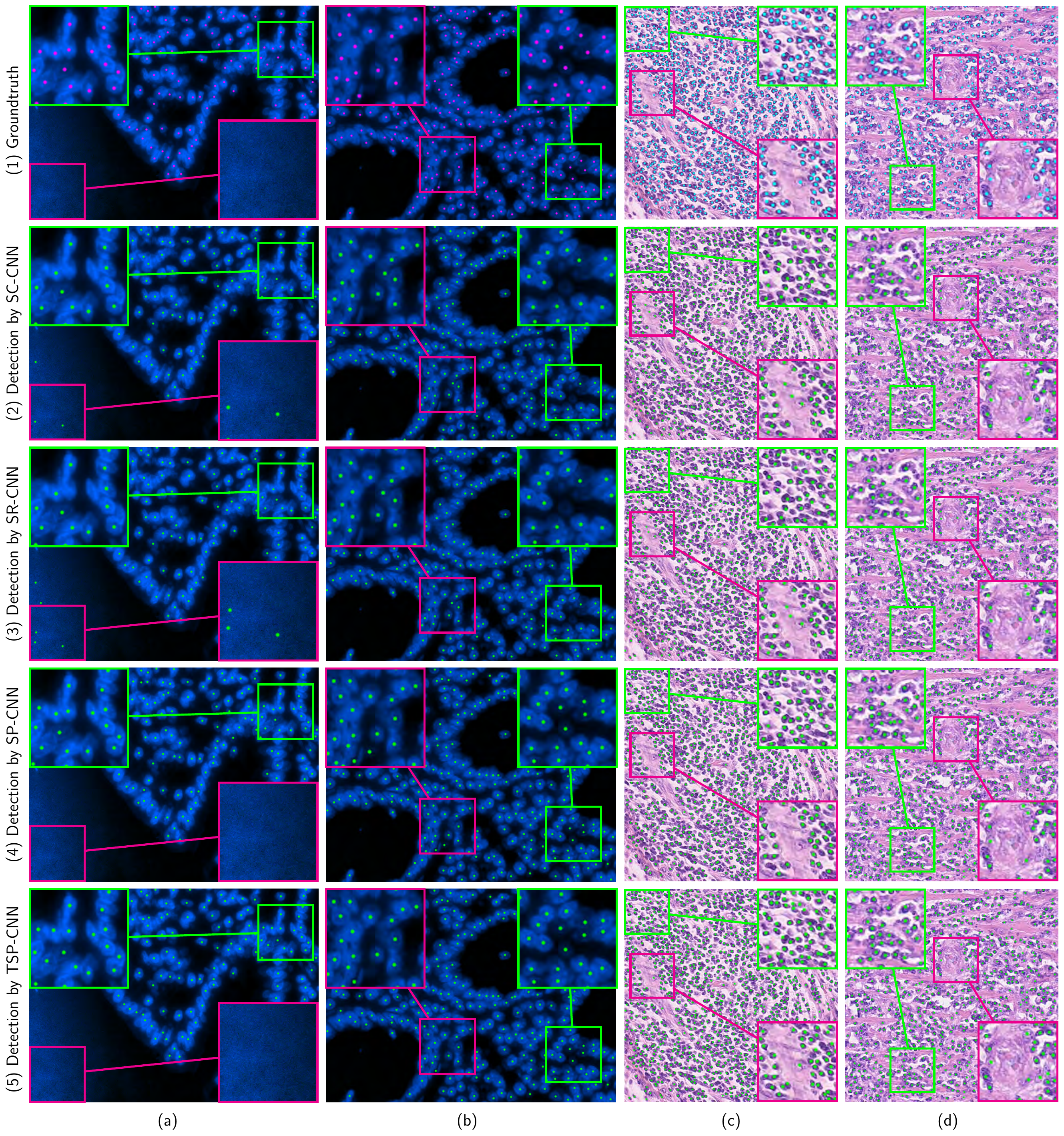}
		\caption{Ground-truth nuclei centers and detection results for two images from each dataset presented here. While `green' zoomed areas show the missed detection by SC-CNN and SR-CNN, `magenta' ones show the wrong detection (FP) of nuclei by those methods. F1-socres for them are as follows: a.2) 0.815, a.3) 0.809, a.4) 0.887, a.5) 0.920; b.2) 0.906, b.3) 0.895, b.4) 0.912, b.5) 0.926; c.2) 0.830, c.3) 0.801, c.4) 0.908, c.5) 0.948; d.2) 0.901, d.3) 0.898, d.4) 0.909, d.5) 0.949.\vspace{-5pt}}
		\label{fig:PSU_dataset}
	\end{figure*}%
	\begin{figure*}[ht!]
		\centering
		\includegraphics[width=1\textwidth]{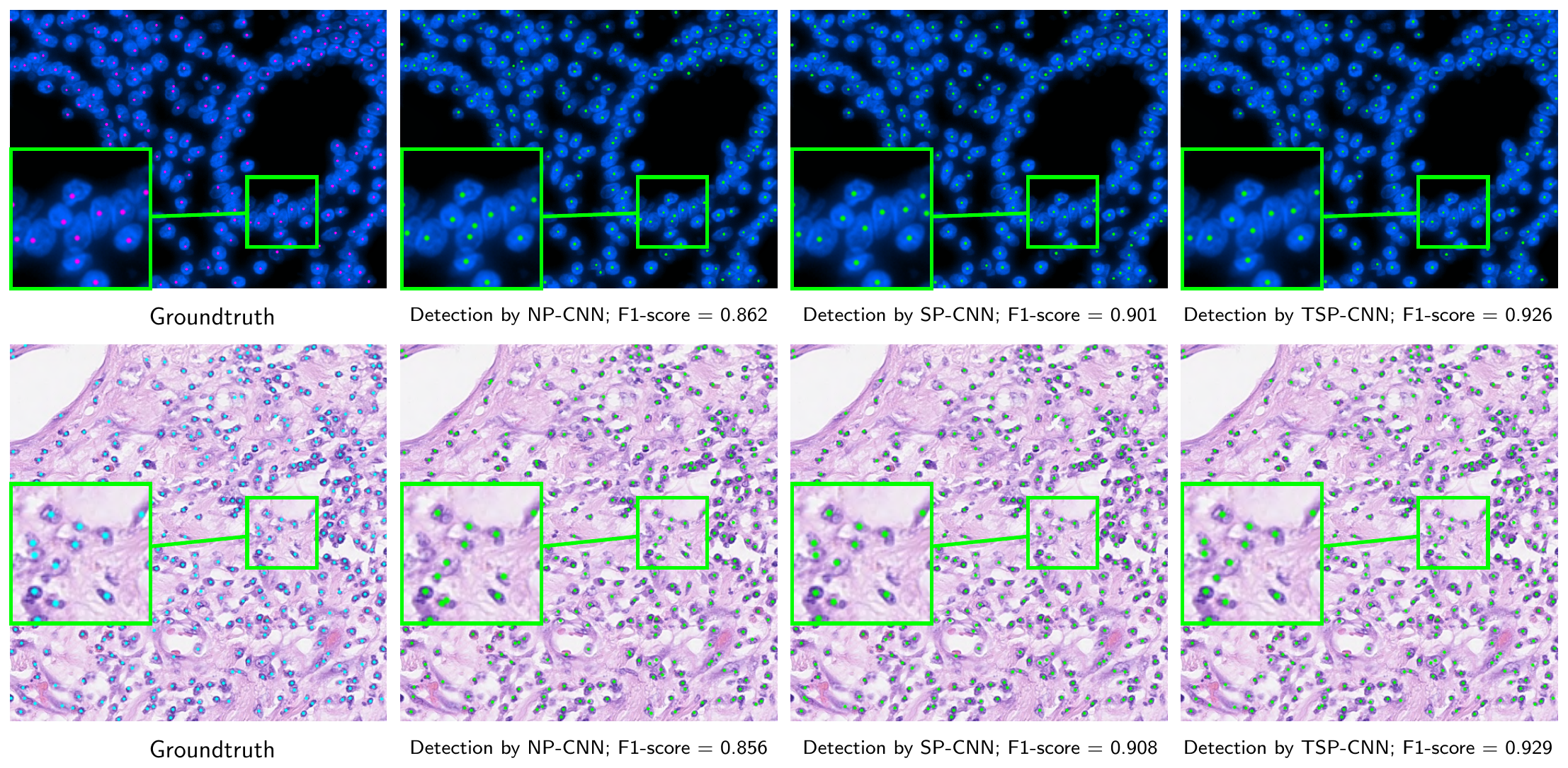}
		% \captionsetup{labelfont={color=red}}
		\caption{Nucleus detection results for TSP-CNN, SP-CNN ($\gamma = 0$),  and NP-CNN ($\gamma = \lambda = 0$) for an example image from the PSU Dataset as well as the UW dataset. The value of prior guided regularization is evident in that false positives are significantly reduced in both SP and TSP-CNN. NP-CNN detects two cells in a region with only one cell (PSU dataset image), and in the example image from UW dataset, there are two missed detections plus one non-cell texture is detected as nuclei.\vspace{-5pt}}
		\label{fig:NP}
	\end{figure*}
	\subsection{Comparison With State-of-the-Art}
	\label{sec:expD}
	As is common, the threshold value $T$ is varied to generate a Precision-Recall curve. The Precision-Recall curve for averaged values over all images in the test set from UW Dataset \cite{SC_CNN} is plotted in Fig. \ref{fig:PRcurve}, and for PSU Dataset the PR-curve is plotted in Fig. \ref{fig:PRcurvePSU}.
	For consistency, results for UW Dataset for the proposed SP-CNN and TSP-CNN are based on using the same assessment procedure as in \cite{SC_CNN}, which used a $50\text{-}50$ split of training vs. test images (using the official assessment source codes of \cite{SC_CNN} provided by the paper's author). For PSU Dataset, a $100\text{-}20$ split of training vs. test images was employed by all competing methods. Fig. \ref{fig:PRcurve} essentially compares SP-CNN and TSP-CNN against state of the art deep learning methods: 1.) SC-CNN \cite{SC_CNN}, 2.) SR-CNN \cite{Xie2015}, 3.) SSAE \cite{SSAE}, 4.) CSP-CNN \cite{CSP_CNN} and two other popular feature and morphology based methods 5.) LIPSyM \cite{LIPSyM}, and 6.) CRImage \cite{CRImage}. 
	
	In the case of UW Dataset, $P, R,$ and $F1$ results for the aforementioned competing methods are obtained directly from comparisons already reported in \cite{SC_CNN,CSP_CNN}. Figures \ref{fig:PRcurve} and \ref{fig:PRcurvePSU} reveal that SP-CNN and TSP-CNN achieve Precision-Recall curves that outperform state of the art alternatives. 
	To obtain a single representative figure of merit for each method, we chose the threshold value, $T$, which maximizes the F1-score correspondingly for each method.
	These `best F1-scores' for each method are then reported in Table \ref{tab:results} for the UW Dataset \cite{SC_CNN}. In Table \ref{tab:results1}, `best F1-scores' are reported for the PSU Dataset. Note that in Table \ref{tab:results1} and Fig. \ref{fig:PRcurvePSU}, we focus on deep learning methods only because: 1.) they are shown to comfortably outperform traditional feature and morphology based methods and 2.) the particular methods in \cite{LIPSyM,CRImage} employ features that are not really appropriate for the PSU dataset leading to vastly degraded results. Further, comparisons with CSP-CNN \cite{CSP_CNN} are not reported for the PSU dataset because CSP-CNN is highly customized for the UW dataset.
	
	To show the value of prior guided regularization, we also report in Tables \ref{tab:results} and \ref{tab:results1}, the results of our network for the case with $\lambda= \gamma = 0$ in Eq. (\ref{finalCostFunc}). Because no regularizers are involved, we call this --  No Prior CNN (NP-CNN). The case with only $\gamma=0$ will reduce Eq. (\ref{finalCostFunc}) to Eq. (\ref{eq:costFunc})  and hence corresponds to SP-CNN. Consistent with SP-CNN, the aforementioned NP-CNN also uses $D = 7$ layers.  The quantitative gains of using shape priors and learning shapes, i.e. (T)SP-CNN vs. NP-CNN are readily apparent in Tables \ref{tab:results} and \ref{tab:results1}. Similar trends can also be seen in Figs.\ \ref{fig:PRcurve} and \ref{fig:PRcurvePSU} in terms of the benefits of shape priors. 
	
	\tcr{As an alternate strategy to address false positives, we perform an experiment which involves training a (prior-less) network with a weighted cost function. That is, we divide the Mean Square Error (MSE) term  in Eq. (\ref{eq:CNN_func}) into two terms: 1) MSE between ground-truth and false positive detections, and 2) MSE that results from missed detections, i.e the loss from false negative. During training, in every iteration (containing two training pairs: input image and corresponding ground-truth labels), false positives and negatives are determined using the ground-truth and subsequently a weighted MSE is computed with a larger weight assigned to false positives\footnote{\tcr{The best weights for these two terms were found using cross-validation to be $0.7$ and $0.3$, respectively.}} -- we call this design Weighted NP-CNN (WNPCNN). We train WNP-CNN with exactly the same network setting and parameters as NP-CNN. The detection results are presented in Table \ref{tab:results} which shows that WNP-CNN with a more complicated training process can achieve modest gains over NP-CNN but is still outperformed by SP/TSP-CNN.}
			
	Note that the results reported in Table \ref{tab:results} and Fig. \ref{fig:PRcurve} use the same Golden Region specified in Section \ref{sec:assessment}: $6$ pixels for the UW Dataset consistent with existing literature \cite{SC_CNN}. To investigate the effect of Golden Region selection, we report F1 scores generated by selecting different Golden Region sizes for the UW dataset in Fig. \ref{fig:GR}. We select the SC-, CSP-, SP- and TSP-CNN to conduct the experiments as they are the top-4 methods from Table \ref{tab:results}. From Fig. \ref{fig:GR}, as expected, F1 score increases with an increase in the golden region size. However, for a fixed Golden Region size, it is the relative performance of different methods that matters and as is evident from Fig. \ref{fig:GR}, this trend remains unchanged. In fact, with a smaller Golden Region such as $2$, the SC-CNN and CSP-CNN performance degrades heavily even as the SP-CNN and TSP-CNN exhibit a much more graceful decay emphasizing higher spatial accuracy of (T)SP-CNN.
	
	Figure \ref{fig:PSU_dataset} provides further insight into the merits of SP-CNN and TSP-CNN for two test images from each dataset. We compare TSP-CNN and SP-CNN with the top two methods from Table \ref{tab:results} and Table \ref{tab:results1}, i.e.\ SC-CNN \cite{SC_CNN} and SR-CNN\cite{Xie2015}. Two parts of each image in Fig.\ \ref{fig:PSU_dataset} are magnified for convenience. While `green' zoomed areas show the missed detection by SC-CNN and SR-CNN, `magenta' ones show the wrong detection (FP) of nuclei by those methods. Thanks to learning guided by pertinent prior information, SP-CNN and TSP-CNN avoid false positives, which are still detected as nuclei by competing state of the art methods.
	
In the same spirit, we present results for TSP-CNN, SP-CNN, and NP-CNN in Fig. \ref{fig:NP}. In this figure, we can observe that shape priors indeed help to improve the detection results and tunable shapes improves the performance of SP-CNN even further.  In Fig. \ref{fig:NP}, note that the detections made by NP-CNN, SP-CNN and TSP-CNN nearly capture all the nuclei centers for the example image from the PSU dataset. Yet, SP-CNN and TSP-CNN perform better because guided by shape priors, false positives are significantly reduced. For the UW dataset, benefits of SP/TSP-CNN are evident over NP-CNN for which two missed detections and one false positive can be seen in the zoomed part of the image.

	\section{Discussion and Conclusion}
	\label{sec:conclusion}
	
	Our contribution is a prior-guided deep network that can enhance cell nuclei detection significantly. Analytically, we develop methods for tractable incorporation of nuclei shape priors (provided by a domain expert) as regularizers in the learning. We also extend the said framework to allow the shapes to be learned from the same training samples instead of assuming them to be fixed. Experimentally, higher F1 scores as well as superior precision-recall curves are achieved by the proposed prior guided networks. 
	
	In this work, we used SSIM based shape similarity to facilitate learning of meaningful shapes largely for tractability reasons. While there are other known similarity measures that can be better than SSIM for comparing nuclei shapes, they are not differentiable w.r.t the underlying shapes and hence the network parameters. A significant direction for future work is the design of custom, new shape similarity measures that can be effective in comparing nuclei shapes while being analytically tractable for incorporation in deep learning frameworks. Multi-scale extension of our prior guided framework forms another viable future research direction.

	% -------------------------------------------------------------------------
	\begin{spacing}{1}
		\bibliographystyle{IEEEbib}
		\bibliography{strings,refs}
	\end{spacing}
\end{document}